\documentclass{PoS}
\usepackage{graphicx}
\usepackage{xcolor}
\usepackage{amsmath}
\usepackage{amssymb}
\usepackage{cite}

\title{Non-minimally coupled gravity and vacuum stability}

\ShortTitle{Non-minimally coupled gravity and vacuum stability}

\author{\speaker{Olga Czerwi\'nska}\\
        Institute of Theoretical Physics, Faculty of Physics, University of Warsaw, Poland\\
        E-mail: \email{olga.czerwinska@fuw.edu.pl}}
        
\author{Zygmunt Lalak\\
        Institute of Theoretical Physics, Faculty of Physics, University of Warsaw, Poland\\
        E-mail: \email{zygmunt.lalak@fuw.edu.pl}}
        
\author{Marek Lewicki\\
        Institute of Theoretical Physics, Faculty of Physics, University of Warsaw, Poland\\
        ARC Centre of Excellence for Particle Physics at the Terascale (CoEPP) $\&$  CSSM, \\Department of Physics, University of Adelaide, South Australia 5005, Australia\\
        E-mail: \email{marek.lewicki@fuw.edu.pl}}
        
\author{Pawe{\l } Olszewski\\
        Institute of Theoretical Physics, Faculty of Physics, University of Warsaw, Poland\\
        E-mail: \email{pawel.olszewski@fuw.edu.pl}}

\abstract{We investigate the properties of vacuum decay taking into account a non-minimal coupling  to gravity. We extend the standard thin-wall solution to include the non-minimal coupling and verify its validity by comparison with a full numerical study. We also investigate the implications of a~large cosmological constant whose influence on the geometry boosts the tunneling rate. Our analysis shows that the influence of the non-minimal coupling differs significantly between the  cases of Minkowski and  deSitter backgrounds. \\
ADP-17-2/T1008 }

\FullConference{Corfu Summer Institute 2016 "School and Workshops on Elementary Particle Physics and Gravity"\\
		31 August - 23 September, 2016\\
		Corfu, Greece}

\begin{document}

\section{Introduction \label{sec:intro}}

Recent discovery of the Higgs boson initiated the thorough investigation of vacuum stability in the Standard Model. Observed data indicate that the electroweak minimum in the SM effective potential is metastable, so the potential can have a second minimum to which the electroweak vacuum may decay \cite{degrassi,buttazzo}. Numerous modifications of the potential and their features have been investigated: additional higher-dimensional interactions \cite{Lalak:2015usa}, approximate scale-invariance \cite{DiLuzio:2015iua}, gauge dependence \cite{Lalak:2016zlv, Plascencia:2015pga}, relation to primordial black holes \cite{Burda:2016mou}, to name a few.

Also the study of the gravitational impact on the metastability has been studied, mostly following the classic work of Coleman and De Luccia \cite{Coleman:1980aw}. Using thin-wall approximation they showed that the decay of Minkowski vacuum into anti-deSitter one in gravitational background is heavily suppressed. Recent works have been devoted to describe the influence of gravity in more detail, mostly in some well-defined cosmological context \cite{Masoumi:2016pqb,Czerwinska:2015xwa,Loebbert:2015eea}.

Central point of this note is the question of the influence of non-minimal coupling between the scalar field and scalar curvature, $\xi$, on vacuum decay.
This coupling is required for the renormalizability of the scalar field in curved spacetime and it is a crucial feature of the Higgs inflation model that is still allowed by the experimental data \cite{Ade:2015lrj,Bezrukov:2007ep}. So far its impact on the vacuum decay has been investigated in case of the inflationary background \cite{espinosa,Herranen:2014cua,Herranen:2015ima} and in the Standard Model case \cite{Isidori:2007vm,Rajantie:2016hkj}.
In this paper we explore a theory with a single scalar field and a renormalizable potential which may seem simplified but it is dictated by the need to accommodate in a readable manner a wide spectrum of parameters all of which are controlling the influence of gravity. Tunneling both close and far from the thin-wall regime is discussed and we consider different geometries of the true and false vacuum. Our discussion aims to be universal and applicable in the plethora of contexts evoking the quantum tunneling in the presence of gravity like inflation \cite{DiVita:2015bha,Bezrukov:2014ipa} or baryogenesis \cite{Lewicki:2016efe} to name a few.

The outline of the paper is as follows. In Section~\ref{sec:model} we present our Lagrangian describing scalar field non-minimally coupled to gravity, analyse the non-minimal coupling's influence on the thin-wall approximation and also numerically calculate the action for the bounce solution to verify our analytical approximation. We also investigate the influence of a large cosmological constant on the decay of the false vacuum and the connection between tunneling via bubble nucleation and the Hawking-Moss solution. In Section \ref{sec:concl} we present our conclusions and summarize the paper. For more details see \cite{Czerwinska:2016fky}.

\section{Model \label{sec:model}}

\begin{figure}[ht]
\begin{center}
\includegraphics[width=0.55\textwidth]{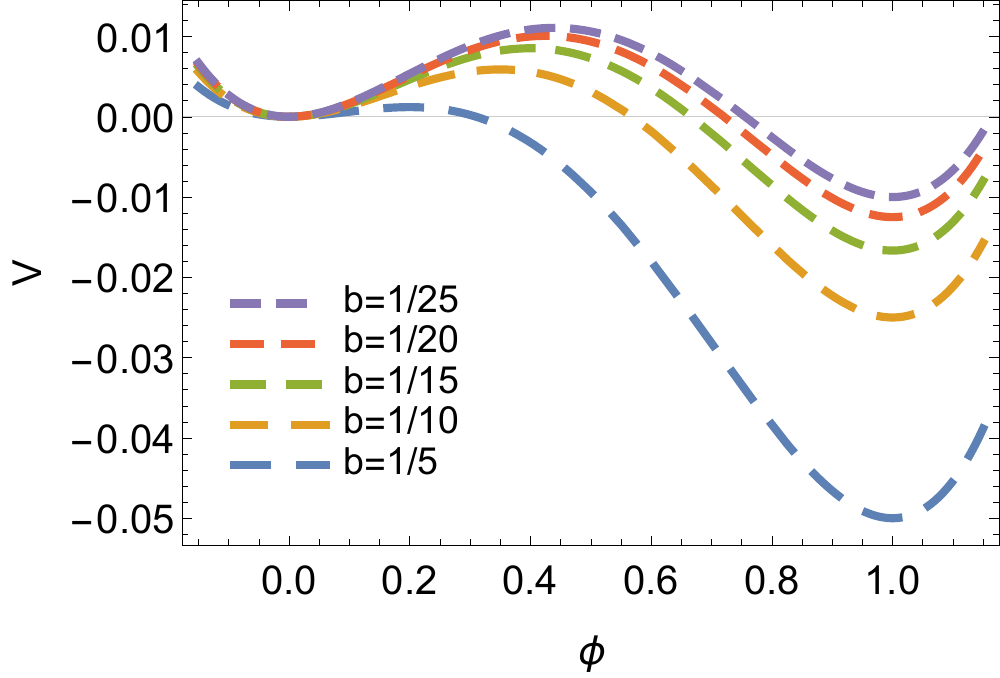}
\caption{
Toy model potential for different values of $b$ parameter and vanishing vacuum energy $c=0$ for $a=1$. \label{fig:potplot}}
\end{center}
\end{figure}

We want to discuss the impact of the gravity on the vacuum decay process considering a very general toy model describing a single neutral scalar field.
Its Lagrangian is suplemented by the two gravitational terms - usual Einstein-Hilbert action and non-minimal coupling of the scalar field to the Ricci scalar:
\begin{equation}
\label{eq:lagrangian}
\mathcal{L}= 
\frac{1}{2}(\partial \phi )^2-V+\frac{1}{2}\frac{R}{\kappa}\left(1-\xi \kappa  \phi^2 \right)
\end{equation}
with the very simple but informative potential 
\begin{equation}
\label{eq:potential}
V= -\frac{1}{4} a^2 (3 b - 1) \phi^2 + \frac{1}{2} a (b - 1) \phi^3 + \frac{1}{4} \phi^4+a^4 c.
\end{equation}
It is intentionally chosen this way as it possesses all features we require to discuss tunneling, the most important is - it has two minima: at $\phi=0$ and $\phi=a$, see Figure \ref{fig:potplot}. We assume that the field is initially in a homogeneous configuration in the more shallow minimum (or {\it false} vacuum) at $\phi=0$ which we will denote by $\phi_{ f}$ and it tunnells to the deeper minimum (or {\it true} vacuum) at $\phi_t=a$.

The only dimensionful parameter is $a$ and the usual choice $a=1$ means that true vacuum is positioned at the Planck scale as we use natural units ($M_p=1$). Decreasing $a$ corresponds to attenuating the gravitational effects and bringing our results closer to flat spacetime case. Constant $c$ is responsible for the character of our initial false vacuum (it corresponds to the vacuum energy) and we focus on a de Sitter false vacuum with $c>0$ and Minkowski false vacuum with $c=0$. Parameter $b$ controls the degeneration of the vacua. Figure~\ref{fig:potplot} depicts our potential in the range of parameters used throughout the paper, we fix $a=1$ and $c=0$.  

\subsection{Tunneling \label{sec:tunnel}}

Our discussion is based on the standard formalism of Coleman and De Luccia (CDL) \cite{Coleman:1980aw}, which assumes that vacuum decay proceeds through nucleation of true vacuum bubbles within our false vacuum. The decay probability is given by \cite{Coleman:1977py,Callan:1977pt}
\begin{equation}\label{eq:tau}
\Gamma =  A e^{-S},
\end{equation} 
where $A$ is a prefactor coming from quantum corrections that is not discussed in the present paper and $S$ is in general the difference of the action integral between final and initial field configurations. For the result including all the gravitational effects these are respectively the Coleman-DeLuccia bounce $\phi_{\mathrm{CDL}}$ and $\phi_{\mathrm{f}}$ (and we denote $S$ by $S_{\mathrm{CDL}}$)
 \begin{equation}\label{eq:CDLaction}
S_{ CDL}=S[\phi_{ CDL}]-S[\phi_f]\;.
\end{equation}

Bubble (bounce) is a spherically symmetric scalar field configuration, $\phi = \phi(\tau)$, fulfilling euclidean EOM with appropriate boundary conditions imposed so that the action difference is finite and coordinate singularities are prevented. Action of such a bubble in presence of non-minimal coupling to gravity reads 
\begin{eqnarray}
\label{eqn:euclideanaction}
& S_E =2\pi^2 \int d \tau \rho^3 \left( \frac{1}{2} \dot{\phi}^2+V-\frac{1}{2} \frac{R}{\kappa}\left(1 -\kappa \xi  \phi^2  \right) \right) \\
& \nonumber = 2\pi^2 \int d \tau \left[ \rho^3 \left( \frac{1}{2} \dot{\phi}^2+V \right)-\frac{3}{\kappa}\left( 1-\xi \kappa \phi^2 \right)
\rho \left(\dot{\rho}^2+1 \right)+6\xi \dot{\phi} \phi \dot{\rho} \rho^2 \right] 
+
\left. \frac{6\pi}{\kappa}\left( 1 - \kappa \xi \phi^2 \right) \rho^2 \dot{\rho} \right|_{0}^{\tau_{ max}} \, 
\end{eqnarray}
with metric $ds^2=d\tau^2 + \rho(\tau)^2(d\Omega)^2$ corresponding to the $FRW$ metric with the curvature parameter $k=+1$. Here $d\Omega$ denotes an infinitesimal element of the $3D$ sphere, $\rho(\tau)$ is the radius of that sphere, $\dot{\phi} = \frac{d\phi}{d\tau}$ and $R=-6\left( \frac{\ddot{\rho}}{\rho}+ \frac{\dot{\rho}^2}{\rho^2} -\frac{1}{\rho^2} \right)$. In case of dS false vacuum the boundary term always vanishes. 

From the above action (\ref{eqn:euclideanaction}) we obtain the equation of motion of the scalar field,
\begin{equation}\label{eq:EFEOM}
 \ddot{\phi}+3\frac{\dot{\rho}}{\rho} \dot{\phi}
-\xi \phi R
 =\frac{\partial V}{\partial \phi}\, ,
\end{equation}
the first 
\begin{equation}\label{eqn:Fried1}
\dot{\rho}^2=
1+ \frac{\kappa \rho^2}{3(1-\kappa \xi \phi^2)}\left(\frac{1}{2}\dot{\phi}^2-V + 6 \xi \dot{\phi}\phi\frac{\dot{\rho}}{\rho}
\right)\, 
\end{equation}
and the second Friedman equation
\begin{equation}\label{eqn:Fried2}
\ddot{\rho}= \frac{\kappa \rho}{ 3\left(1-\kappa \xi \phi^2\right)} 
\left(-\dot{\phi}^2-V+3\xi \left(\dot{\phi}^2+\ddot{\phi}\phi+\dot{\phi}\phi\frac{\dot{\rho}}{\rho} \right)
\right)\, .
\end{equation}
Using the first Friedman equation we can also further simplify the action (\ref{eqn:euclideanaction}) 
\begin{equation}\label{eqn:euclideanactionsimple}
S_E = 4\pi^2 \int d \tau \left[ \rho^3 V -\frac{3\rho}{ \kappa}\left( 1-\xi \kappa \phi^2 \right) \right] +\left. \frac{6\pi}{\kappa}\left( 1 - \kappa \xi \phi^2 \right) \rho^2 \dot{\rho} \right|_{0}^{\tau_{ max}}.
\end{equation}

Scale factor $\rho$ crosses zero at least once \cite{Guth:1982pn} and without loss of generality we can chose value of $\tau$ of the first zero to be $\tau=0$ and the other at $\tau_{\rm max}$. Then the appropriate boundary conditions read
\begin{eqnarray}
\label{eqn:boundaryconditions}
& \dot{\phi}(0) =\dot{\phi}(\tau_{ max})=0,  \nonumber \\
& \rho(0) =0, \nonumber   \\
& \rho(\tau_{ max})  =0 \quad  \quad \quad  {\rm  (for \ dS \ false \ vacuum)}, \nonumber \\
& \rho(\tau_{ max})  =\rho_{ max}\neq 0 \quad \quad {  \rm (for \ Minkowski \ false \ vacuum)} . \nonumber
\end{eqnarray}

In our definition of $R$ second power of $\rho$ appears in the denominator so there is a possibility of divergence here. Therefore, it is much more convenient for numerical calculations to express the scalar curvature using the Friedman equations as
\begin{equation}\label{eqn:R}
R=-6\left( \frac{\ddot{\rho}\rho+\dot{\rho}^2-1}{\rho^2} \right)
=
 \frac{\kappa}{ \left(1-\kappa \xi \phi^2\right)} 
\left(\dot{\phi}^2+4V-6\xi \left(\dot{\phi}^2+\phi\ddot{\phi}
+3\dot{\phi}\phi\frac{\dot{\rho}}{\rho}\right)
\right).
\end{equation}
This way $R$ contains only the Hubble parameter that is already present in the scalar field's EOM and thus has to be numerically stable.

\begin{figure}[ht]
\begin{center}
\includegraphics[height=4.6cm]{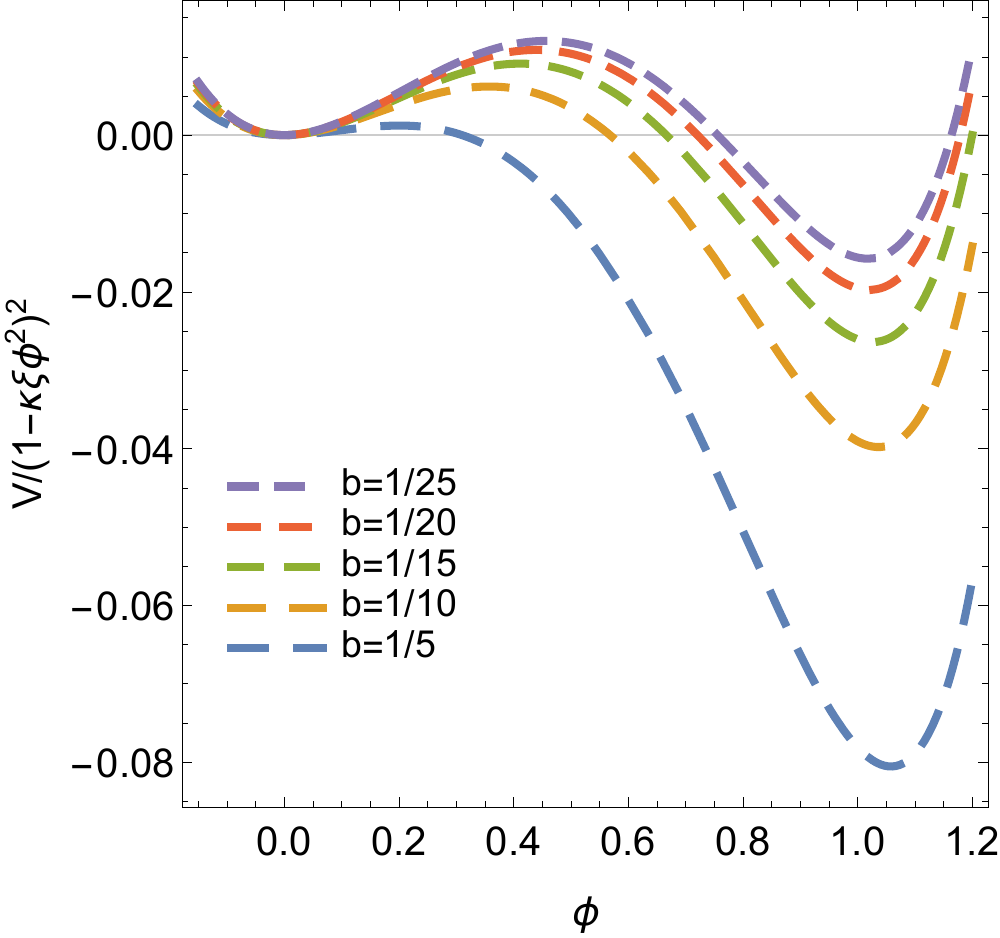}
\includegraphics[height=4.6cm]{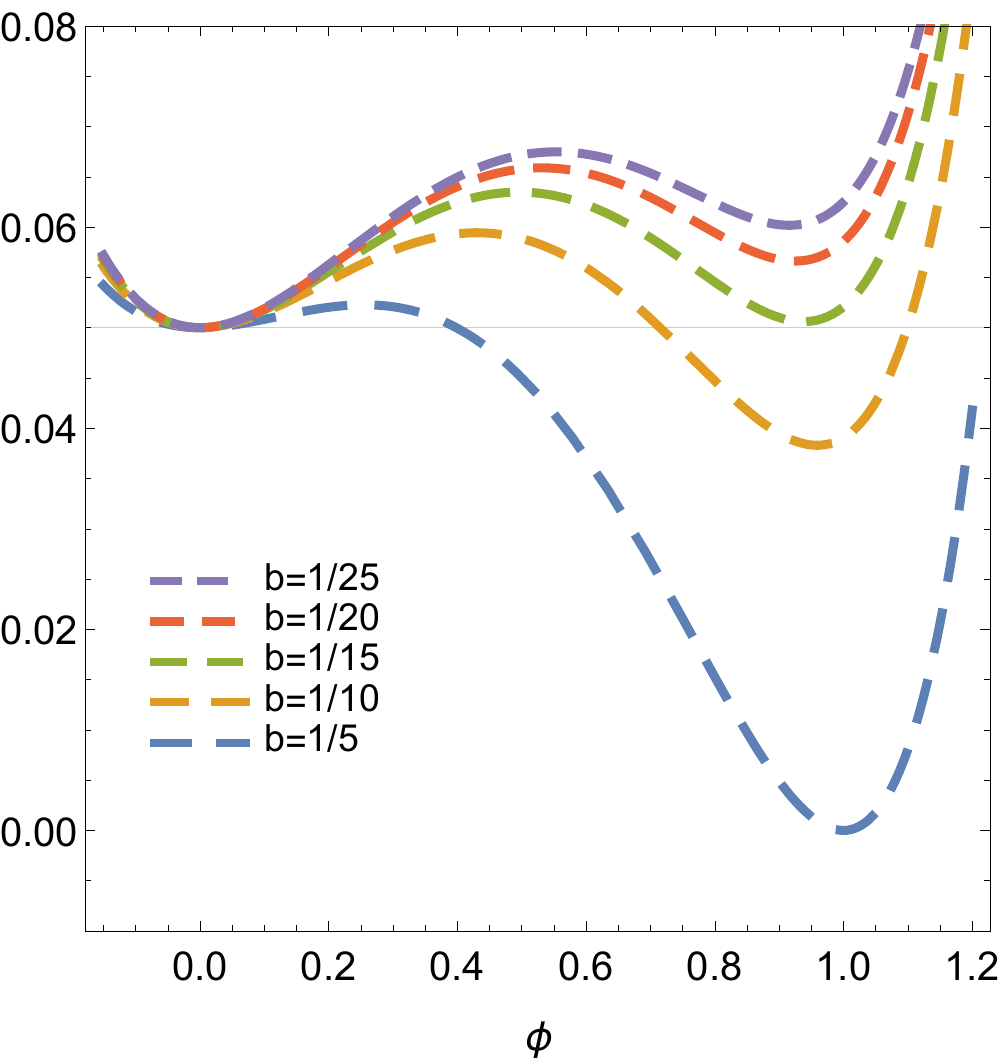}
\includegraphics[height=4.6cm]{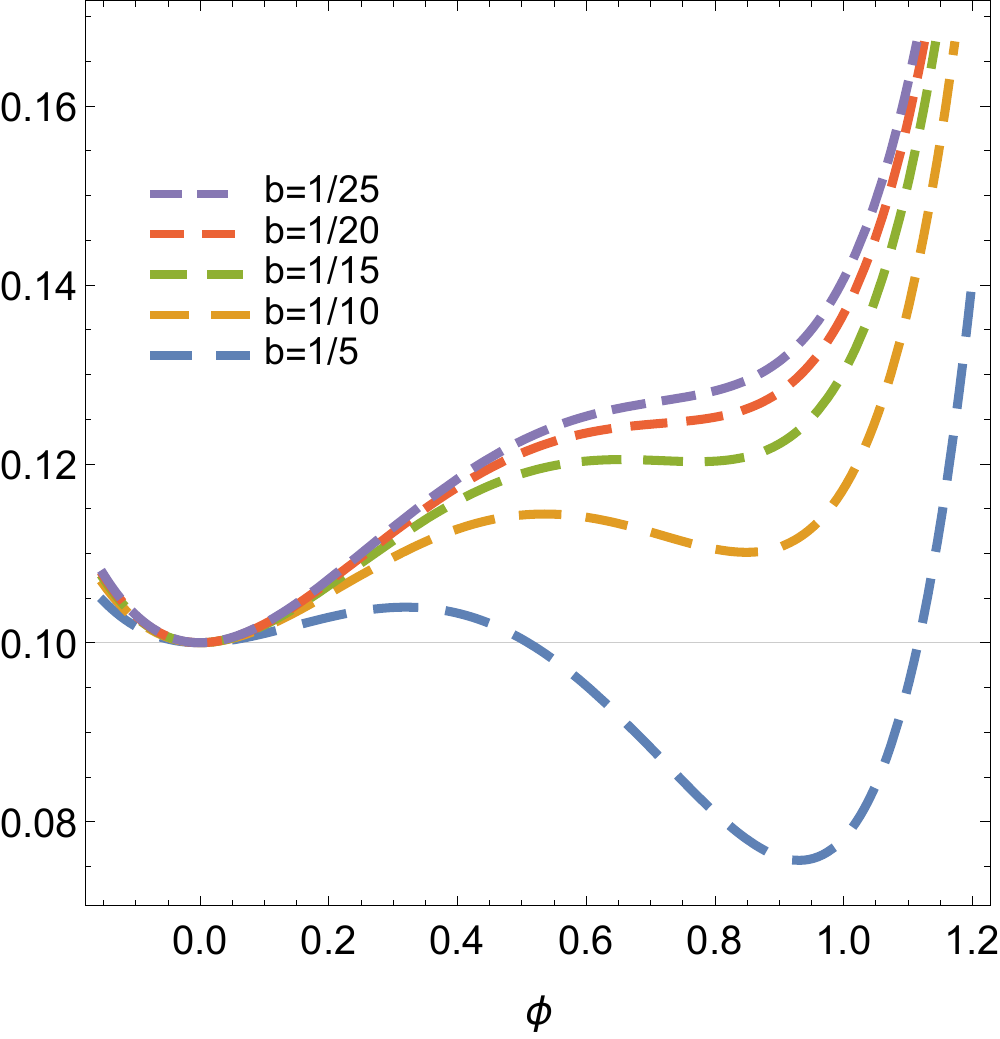}
\caption{ Modified potential $V/(1-\kappa \xi \phi^2)^2$ for different choices of the vacuum energy $c$ and with the non-minimal coupling set to $\xi=0.2$. Value of the c constant was set to c = (0; 0:05; 0:1) from left to right. \label{fig:modpotplot}}
\end{center}
\end{figure}

For gravitational background we assume a constant field configuration, which results in the simplified first Friedmann equation (\ref{eqn:Fried1})
\begin{equation}
\frac{d \rho}{d \tau}=\sqrt{1-\frac{\kappa \rho^2 V}{3\left(1-\kappa \xi \phi^2\right)}}\, ,
\end{equation}
where $V=V(\phi)$ and $\phi$ is the chosen constant field value. Equation (\ref{eqn:euclideanactionsimple}) corresponds then to  
\begin{eqnarray}\label{eqn:backgroundaction}
& S[\phi_{\rm f}] =-\frac{24 \pi^2 (1- \kappa \xi \phi_{\rm f}^2)^2}{\kappa^2 V_{\rm f}}
 \quad \  ({\rm for \ dS}), \\
& S[\phi_{\rm f}] =0  \quad \quad \quad ({\rm for \ Minkowski})\, .
\end{eqnarray}
In our potential false vacuum is always positioned at $\phi_{\rm f} =0$ so there is no modification of the false vacuum energy. However, the same reasoning applies to the true vacuum energy which is modified by the non-minimal coupling - energy of true vacuum can be increased beyond that of the false vacuum (in the case when $V(\phi_{\mathrm{t}})>0$) making false vacuum stable. This is especially visible for large vacuum energies where the true vacuum can disappear altogether as shown in Figure~\ref{fig:modpotplot}. We always neglect tunneling in such cases because even though the bubble profile can sometimes still be calculated, such bubble is not energetically favourable and would not grow after nucleation.

\subsection{TW approximation \label{sec:tw}}

Thin-wall (TW) approximation originating from \cite{Coleman:1980aw,Coleman:1977py}, assumes the true vacuum bubble stretches to some $\bar{\rho}$ having a constant value $V_{\rm t}$ inside and on the outside of the bubble our solution is identical to the false vacuum $V_{\rm f}$. The overall action
\begin{equation}
S_{TW} = S_{\text{wall}}  + S [\phi_{\text{tv}}] - S [\phi_{\text{fv}}]
\end{equation}
acquires the new term $S_{\text{wall}}$ describing the wall. One of the advantages of this approximation is the fact that in some cases it can provide the analytical expression for the bounce solution.

We develop the original setup by including the non-minimal coupling to gravity. The full EOM reads then
\begin{equation}\label{eq:simpEFEOM}
 \ddot{\phi} + \frac{3}{\tau} \dot{\phi}
-\xi \phi R
 =\frac{\partial V}{\partial \phi}\, ,
\end{equation}
where according to TW approximation we neglect the term proportional to $\dot{\phi}$ for the wall and $\rho = \bar{\rho} = \text{const}$ so that curvature can be approximated by $R\approx\frac{6}{\bar{\rho}^2}$.
Integrating (\ref{eq:simpEFEOM}) once we obtain 
\begin{equation}
\frac{d \phi}{d \tau}=-\sqrt{2(V-V_{\rm t})+\xi R \left(\phi^2 -\phi_{\rm t}^2\right)},
\end{equation} 
so the action of the bubble wall reads
\begin{eqnarray}\label{eqn:bubblewallexpansion}
& B_{\rm wall} = 2 \pi^2 \bar{\rho}^3 \int_{0}^{\tau_{\rm{max}}} \left[2 (V - V_{\rm t}) + \xi R (\phi^2 - \phi_{\rm t}^2)\right] d \tau \\
 & \approx 2 \pi^2 \left( \bar{\rho}^3 \int_{\phi_{\rm f}}^{\phi_{\rm t}}
\sqrt{2 (V - V_{\rm t})} d \phi
+
\xi  \bar{\rho}\int_{\phi_{\rm f}}^{\phi_{\rm t}} 
\frac{3\left(\phi^2-\phi_{\rm t}^2\right)}{\sqrt{2 (V - V_{\rm t})}}  d \phi \right) \\
&=2\pi^2 \left( \bar{\rho}^3 S_0+\xi \bar{\rho}S_1 \right),
\end{eqnarray}
where we expanded to the first order in $\xi$. $S_0$ is the usual result we would obtain neglecting gravity: $S_0=\rho_0 (V_{\rm f}-V_{\rm t})/3$, where $\rho_0$ is the size of the bounce obtained numerically neglecting gravity, and $S_1$ is the linear correction due to the non-minimal coupling.

Expansion to the second order in $\xi$ slightly increases the action but it does not improve significantly the accuracy of our results. We prove in Section \ref{sec:numer} that both approximations we use overestimate the correct result. Interpretation of this error presumes that it comes from our assumption on the shape of the bounce rather than from the expansion in the non-minimal coupling $\xi$.

Gravitational part of the action corresponds again to the constant field configuration in (\ref{eqn:euclideanactionsimple}) with some upper limit of the integratiom, namely the radius of the bubble $\bar{\rho}$, which results in
\begin{equation}
S_{\rm grav}=2\pi^2 \frac{2}{3} \frac{\left(1-\bar{\rho}^2 \Lambda V\right)^{3/2} -1 }{\Lambda^2 V } \, ,
\end{equation}
where $\Lambda=\kappa/(1-\kappa \xi \phi^2)$ and $\phi$ is the constant field value.

Final expression for the action for TW approximation including gravitational contributions reads
\begin{equation}\label{eqn:thinwallaction}
S_{ TW}= 2 \pi^2 \left( \bar{\rho}^3 S_0 + \xi \bar{\rho} S_1 - \frac{2}{3} \frac{(1 - \bar{\rho}^2 \Lambda_{\text{f}} V_{\text{f}} )^{3/2} - 1}{\Lambda_{\text{f}}^2 V_{\text{f}}} + \frac{2}{3} \frac{(1 - \bar{\rho}^2 \Lambda_{\text{t}} V_{\text{t}} )^{3/2} - 1}{\Lambda_{\text{t}}^2 V_{\text{t}}} \right),
\end{equation}
where $\Lambda_{\rm f}=\kappa/(1-\kappa \xi \phi_{\rm f}^2)$ and $\Lambda_{\rm t}=1/(1-\kappa \xi \phi_{\rm t}^2)$ are constant field values. 
In the flat case - Minkowski background with $V_{ f}=0$, false vacuum gravity action should be replaced with the appropriate limit $S_{ grav}\rightarrow{V \rightarrow 0}-2\bar{\rho}^2 / \Lambda$ which gives the overall action of the form
\begin{equation}\label{eqn:thinwallactionminkowski}
S_{ TW}= 2 \pi^2 \left( \bar{\rho}^3 S_0 + \xi \bar{\rho} S_1 + \frac{\bar{\rho}^2}{\Lambda_{\text{f}}} + \frac{2}{3} \frac{(1 - \bar{\rho}^2 \Lambda_{\text{t}} V_{\text{t}} )^{3/2} - 1}{\Lambda_{\text{t}}^2 V_{\text{t}}} \right).
\end{equation}

Bi-quadratic equation for the size of the bubble $\bar{\rho}$ used in (\ref{eqn:thinwallaction}) or (\ref{eqn:thinwallactionminkowski}) can be obtained by differentiating the action with respect to $\bar{\rho}$ and  expanding it to the linear order in $\xi$ reads
\begin{eqnarray}\label{eqn:twbubblesize}
& \left[ \left( \frac{1}{\Lambda_{\text{f}}^2} - \frac{1}{\Lambda_{\text{t}}^2} \right)^2 - 3 \xi S_0 S_1 \left( \frac{1}{\Lambda_{\text{f}}^2} + \frac{1}{\Lambda_{\text{t}}^2} \right) \right] + \bar{\rho}^4 \left[ \frac{9}{2} S_0^2 \left( \frac{V_{\text{f}}}{\Lambda_{\text{f}}} + \frac{V_{\text{t}}}{\Lambda_{\text{t}}} \right) + \left( \frac{V_{\text{t}}}{\Lambda_{\text{t}}} - \frac{V_{\text{f}}}{\Lambda_{\text{f}}} \right)^2 + \frac{81}{16} S_0^2 \right] + \\
\nonumber & + \bar{\rho}^2 \left[ -2 \left( \frac{V_{\text{f}}}{\Lambda_{\text{f}}^3} + \frac{V_{\text{t}}}{\Lambda_{\text{t}}^3} \right) - \frac{9}{2} S_0^2 \left( \frac{1}{\Lambda_{\text{f}}^2} + \frac{1}{\Lambda_{\text{t}}^2} \right) + \frac{2}{\Lambda_{\text{f}} \Lambda_{\text{t}}} \left( \frac{V_{\text{f}}}{\Lambda_{\text{t}}} + \frac{V_{\text{t}}}{\Lambda_{\text{f}}} \right) + 3 \xi S_0 S_1 \left( \frac{V_{\text{f}}}{\Lambda_{\text{f}}} + \frac{V_{\text{t}}}{\Lambda_{\text{t}}} + \frac{9}{4} S_0^2 \right) \right] = 0.
\end{eqnarray}

In the flat case our equation of motion for the scalar field simplifies to 
{\begin{equation}\label{eq:FlatEOM}
 \ddot{\phi}+\frac{3}{\tau} \dot{\phi}
 =\frac{\partial V}{\partial \phi}\, 
\end{equation}}
with new boundary conditions
\begin{eqnarray}
& \dot{\phi}(0)=\dot{\phi}(\tau_{\rm max})=0, \nonumber \\
& \lim \limits_{\tau \to \infty} \phi = V_{\rm f}   \, .
 \nonumber
\end{eqnarray}

\subsection{HM solution \label{sec:hm}}

\begin{figure}
\centering
\includegraphics[width=0.4\textwidth]{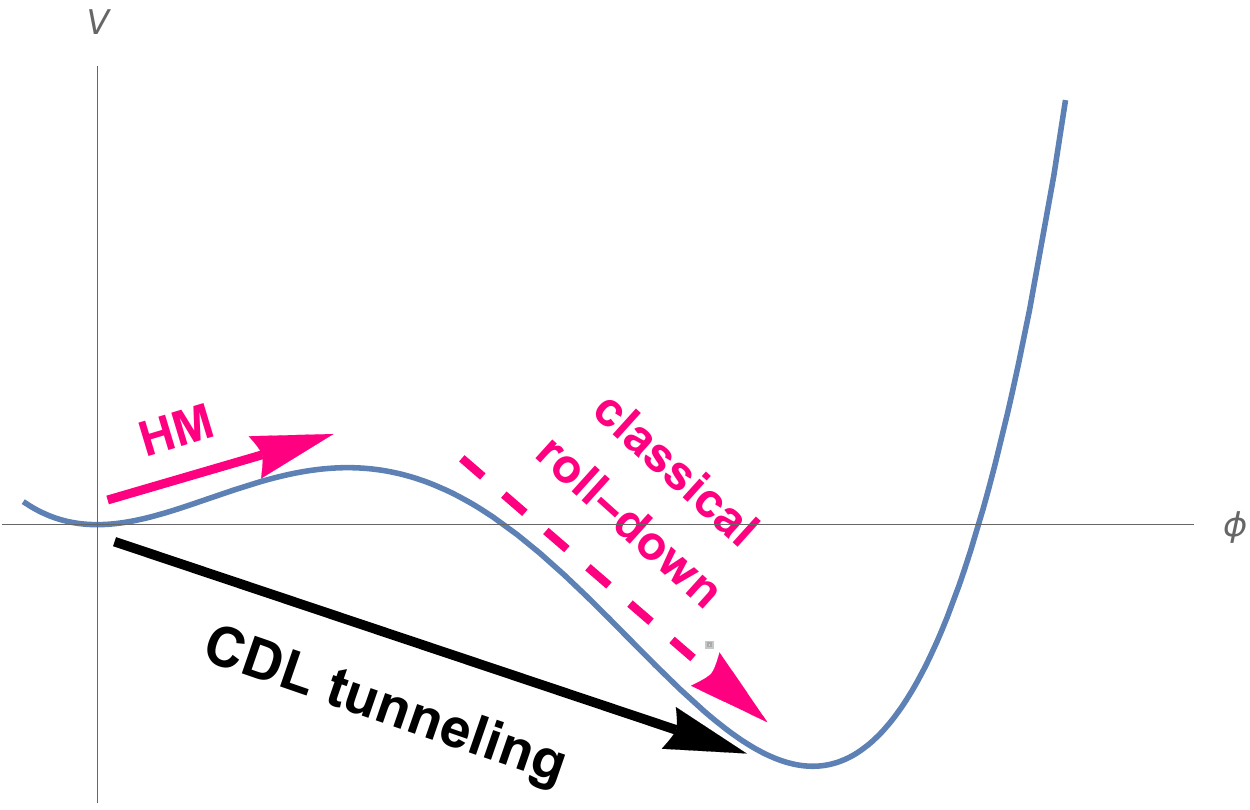}
\caption{Comparison between Coleman-deLucia and Hawking-Moss solutions. \label{fig:hm}}
\end{figure}

Hawking-Moss (HM) instantons describe the probability for a whole horizon volume to transition to the top of the barrier and continue by a classical roll-down \cite{Hawking:1981fz}, see Figure \ref{fig:hm} for comparison with CDL instanton. Action of HM instanton is the difference between action of our false vacuum and the energy of a homogenous solution on top of the potential barrier. Non minimal coupling modifies these energies as described in (\ref{eqn:backgroundaction}) and we get 
\begin{equation}
S_{\rm HM} = - \frac{24 \pi^2 (1- \kappa \xi \phi_{\rm max}^2)^2}{\kappa^2 V_{\rm max}} +
\frac{24 \pi^2 (1- \kappa \xi \phi_{\rm f}^2)^2}{\kappa^2 V_{\rm f}},
\end{equation}
where $\phi_{\rm max}$ and $V_{\rm max}$ correspond to the potential and field values at the top of the barrier.

\subsection{Numerical results \label{sec:numer}}

In a flat case we solve equation of motion for the scalar field (\ref{eq:FlatEOM}) numerically using the shooting method similar to \cite{Lalak:2014qua} and we find the bubble size $\rho_0=\tau \left(\phi=\frac{V_{\rm t}+V_{\rm f}}{2} \right)$ later used in (\ref{eqn:thinwallaction}). This method of obtaining the bubble size is more accurate than the simpler flat spacetimete thin-wall result so the initial flat spacetime error does not influence our thin-wall approximation concernig gravity.

\begin{figure}[ht]
\begin{center}
\includegraphics[width=0.32\textwidth]{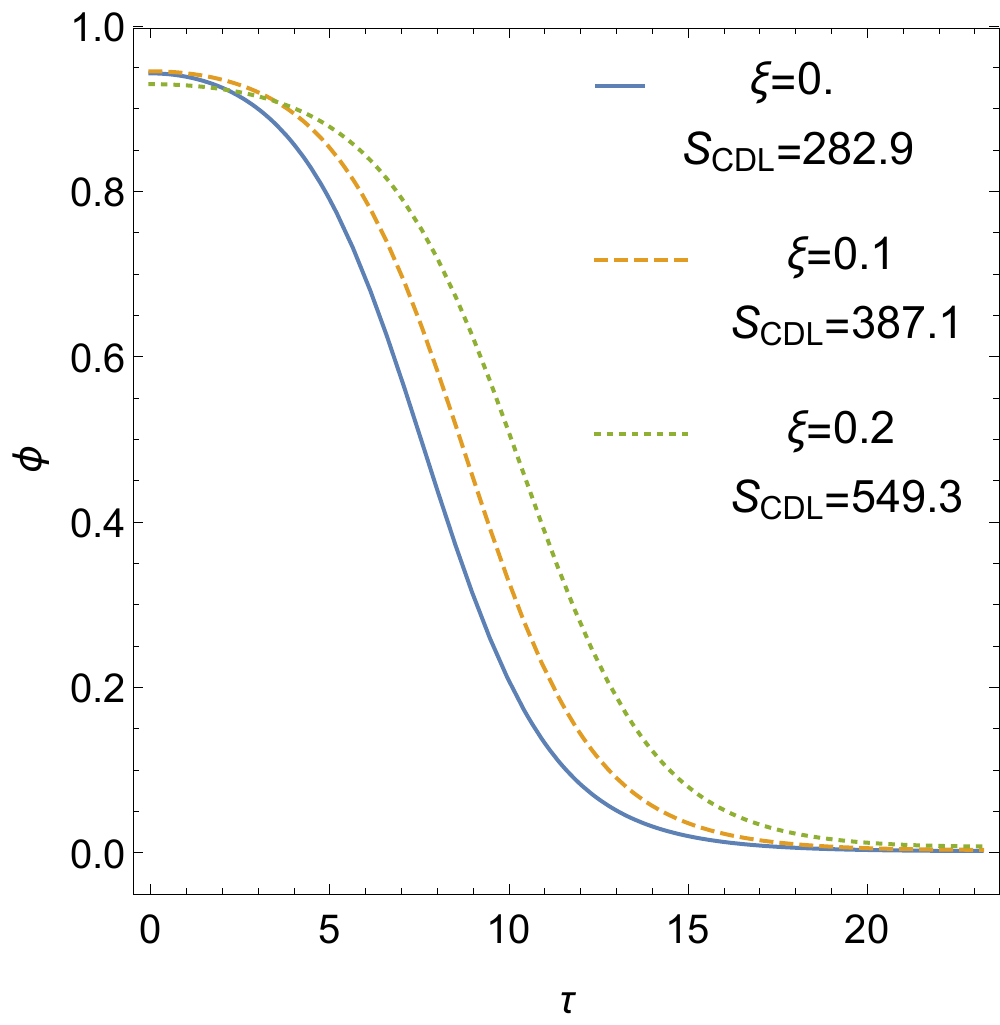}
\includegraphics[width=0.32\textwidth]{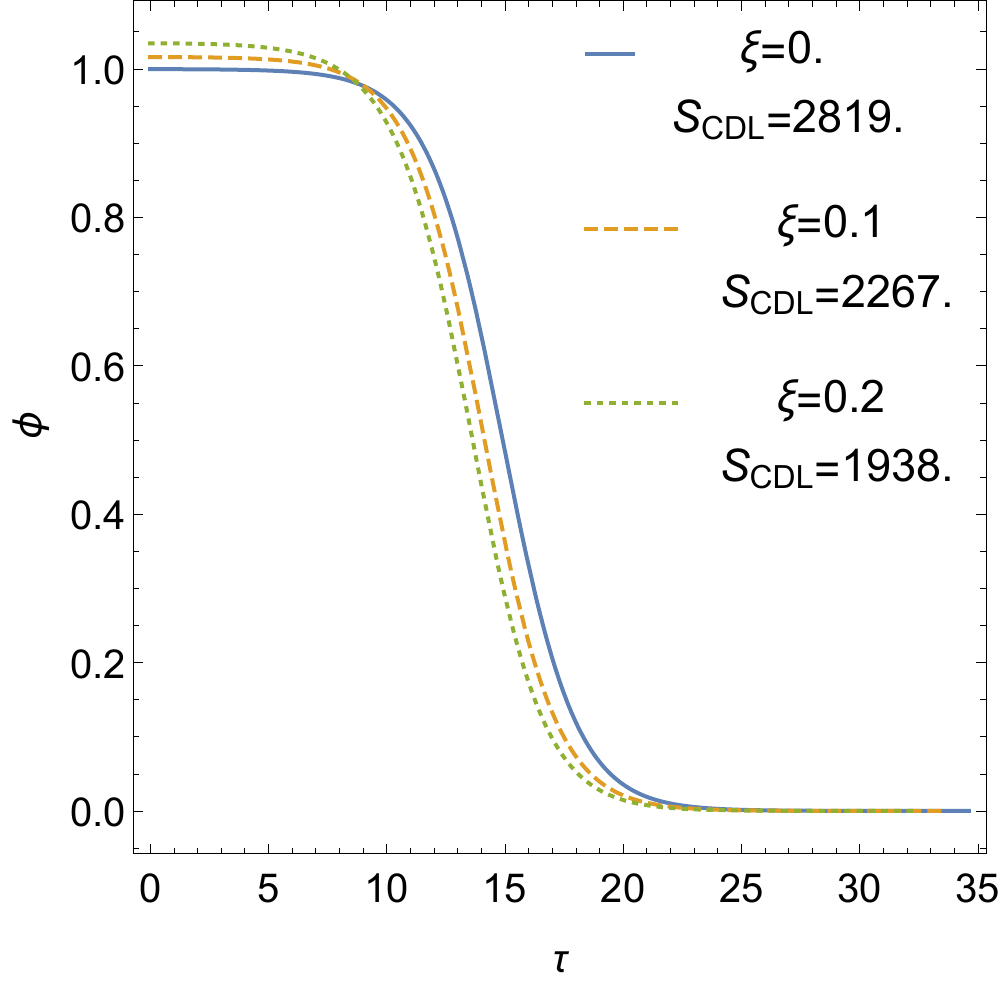}
\caption{ CDL bubble profiles for tunneling from dS false vacuum $c=0.05$ (left) and from Minkowski false vacuum (right) for several values of the non-minimal coupling $\xi$ for $b=1/10$. \label{fig:instantons}}
\end{center}
\end{figure}

In numerical calculations containing $\xi$ we solve the field EOM (\ref{eq:EFEOM}) with Ricci scalar expressed by the scalar field (\ref{eqn:R}) and the second Friedman equation (\ref{eqn:Fried2}) with boundary conditions (\ref{eqn:boundaryconditions}), approximating $\rho(0)$ as proportional to initial $\tau = \epsilon$ and $\dot{\rho}=1$. Our EOM is the equation of a particle in potential $-V(\phi)$ with a time-dependent friction $3 \frac{\dot{\rho}}{\rho}$ and we expand it around both vacua in $\epsilon$ that can be arbitrarily small. We neglect higher orders of both expansions and we find iteratively their leading orders that fulfill the boundary condition. Last initial condition - field value $\phi_0$ corresponding to CDL, can be found by a undershoot/overshoot method known from the flat setup, see e.g. \cite{Lalak:2015usa}. Resulting bubble profiles are presented in Figure~\ref{fig:instantons}.

\begin{figure}[ht]
\begin{center}
\includegraphics[width=0.32\textwidth]{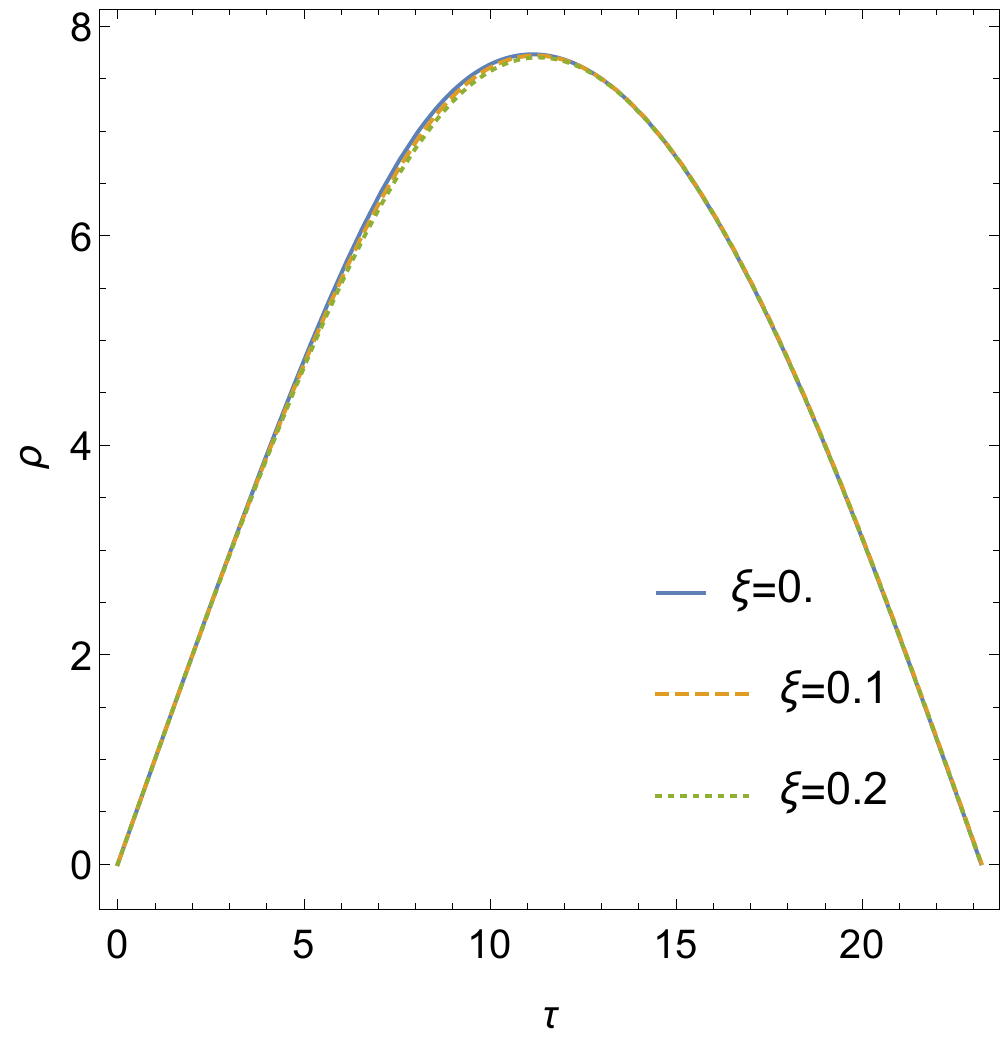}
\includegraphics[width=0.32\textwidth]{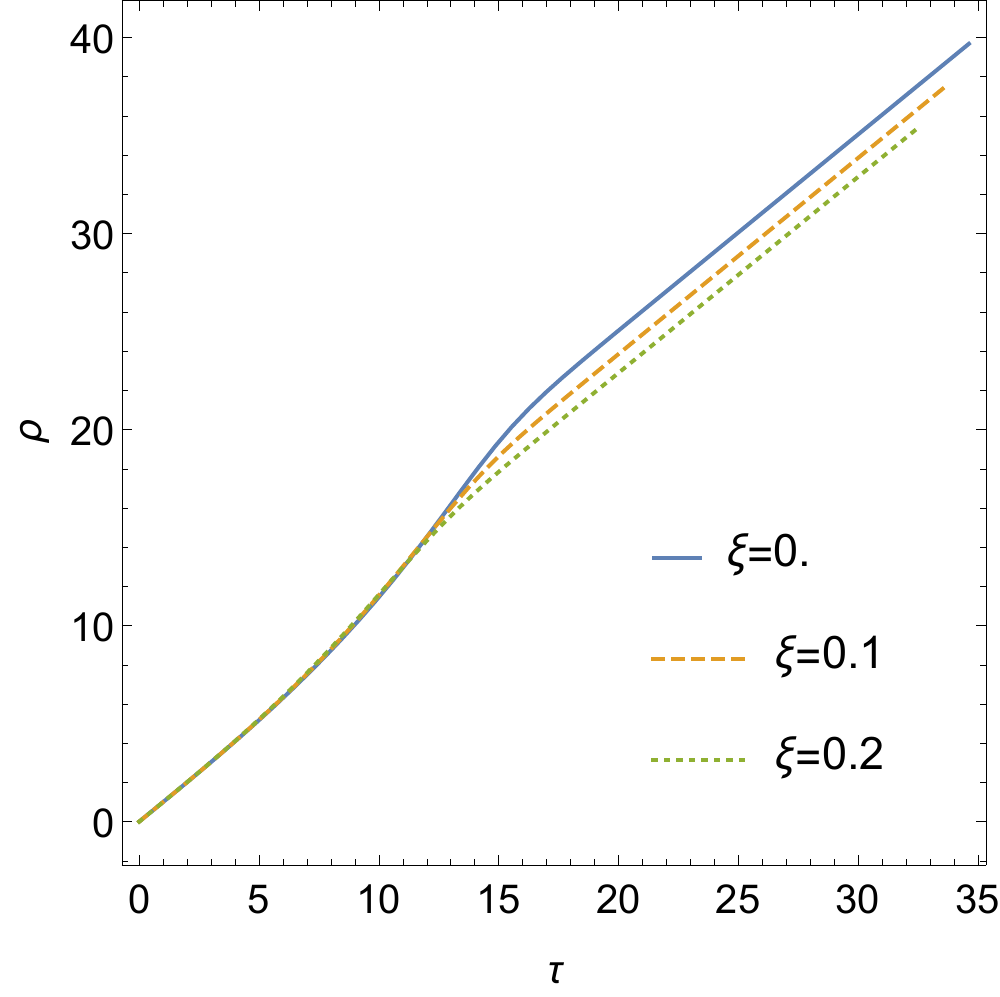}
\caption{ Bubble radius for tunneling from dS false vacuum $c=0.05$ (left panel) and from Minkowski false vacuum (right Panel) for several values of the non-minimal coupling $\xi$ for $b=1/10$. \label{fig:scale}}
\end{center}
\end{figure}

Now also the influence of the boundary term in the action (\ref{eqn:euclideanactionsimple}) when the false vacuum has a vanishing energy is straightforwardly visible. We can see in the the Figure \ref{fig:scale} that $\rho$ asymptotes to a linear function instead of crossing zero again at $\tau_{\rm max}$ and the effect of the boundary term is not negligible. 

\begin{figure}[ht]
\begin{center}
\includegraphics[width=0.32\textwidth]{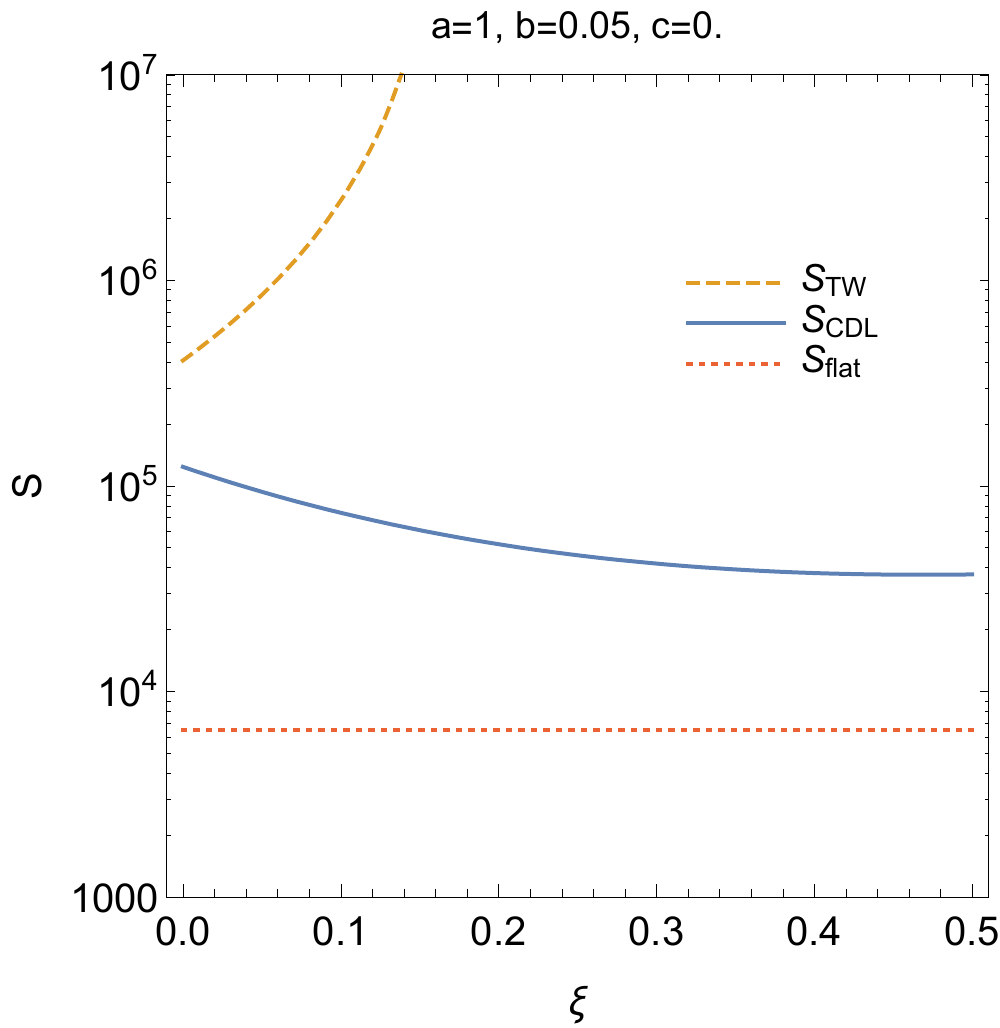}
\includegraphics[width=0.32\textwidth]{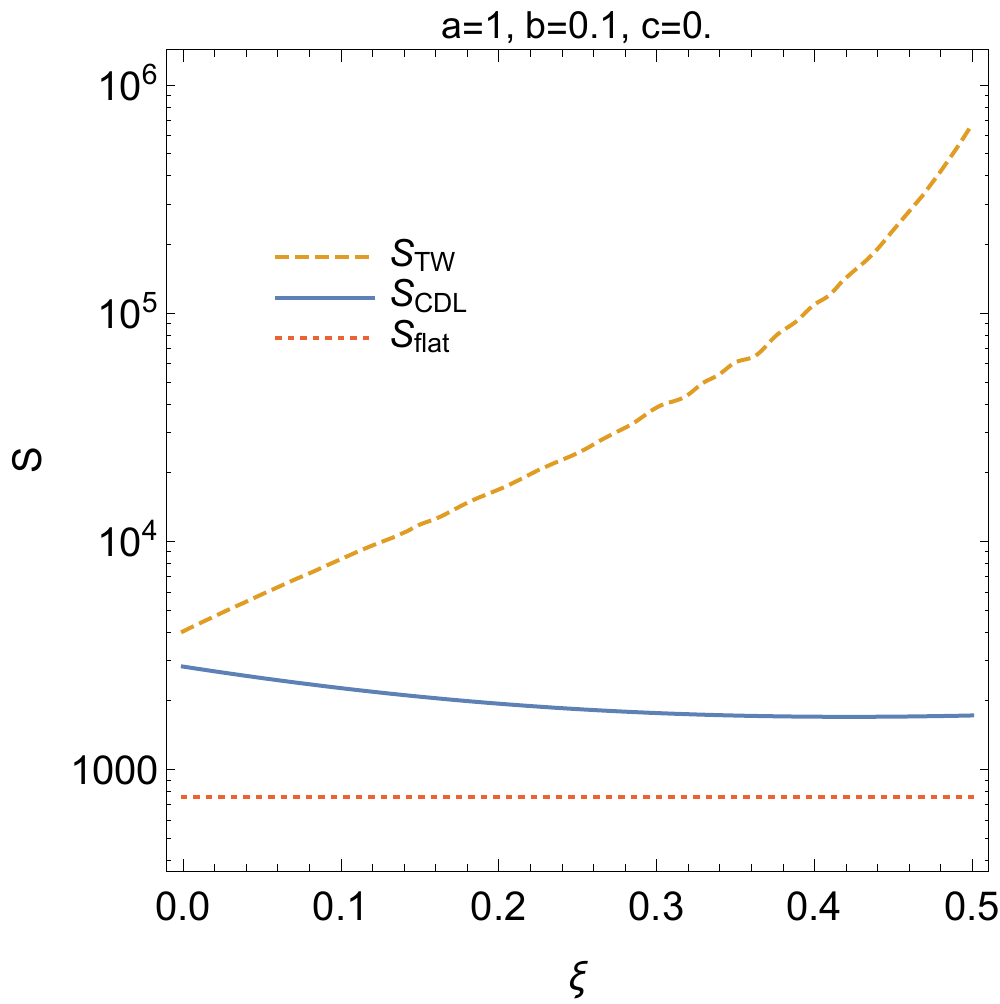}
\caption{ Tunneling action for Minkowski false vacuum as a function of non-minimal coupling obtained using four different methods for different values of $b$ parameter. $S_{\rm CDL}$ is the numerically obtained result fully including gravity,  $S_{\rm TW}$ is the result of our thin-wall approximation, $S_{\rm HM}$ comes from Hawking-Moss solution and $S_{\rm flat}$ is the, numerically obtained, flat spacetime result completely neglecting gravity. 
 \label{fig:actionMin}}
\end{center}
\end{figure}

\begin{figure}[ht]
\begin{center}
\includegraphics[width=0.32\textwidth]{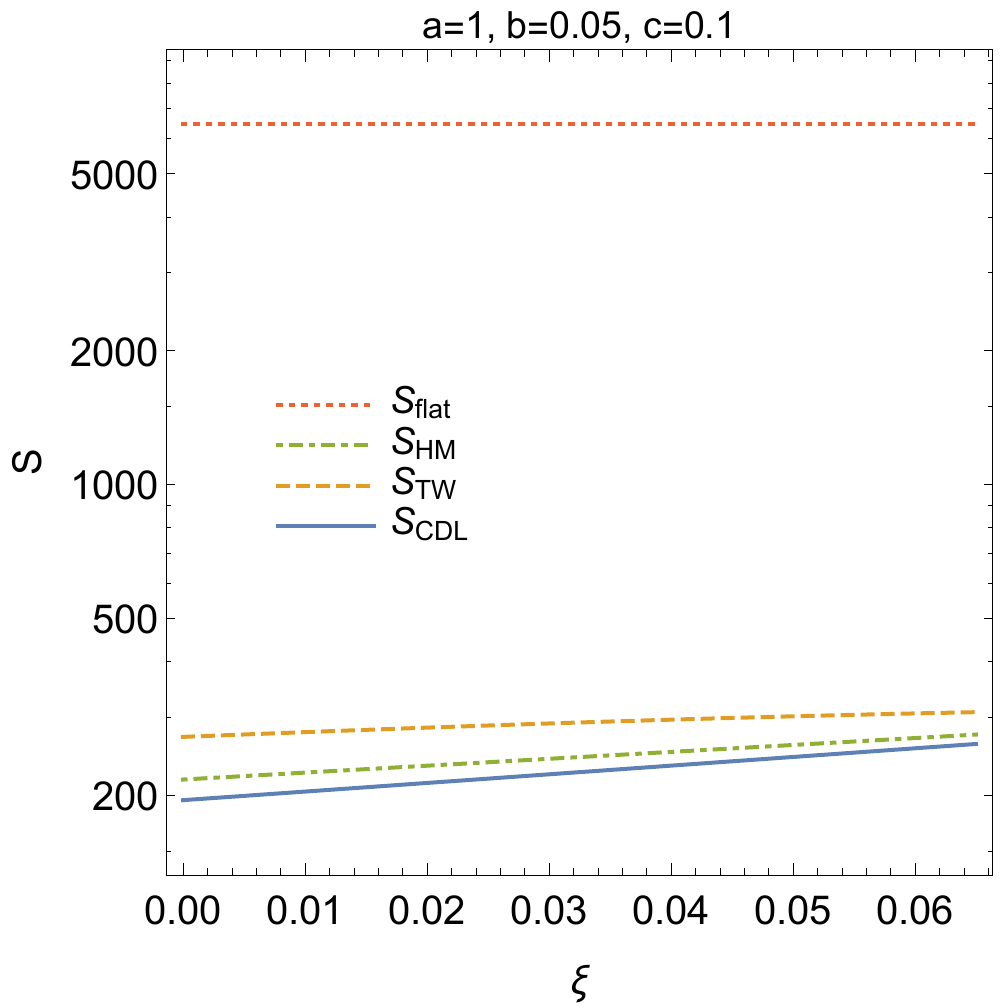}
\includegraphics[width=0.32\textwidth]{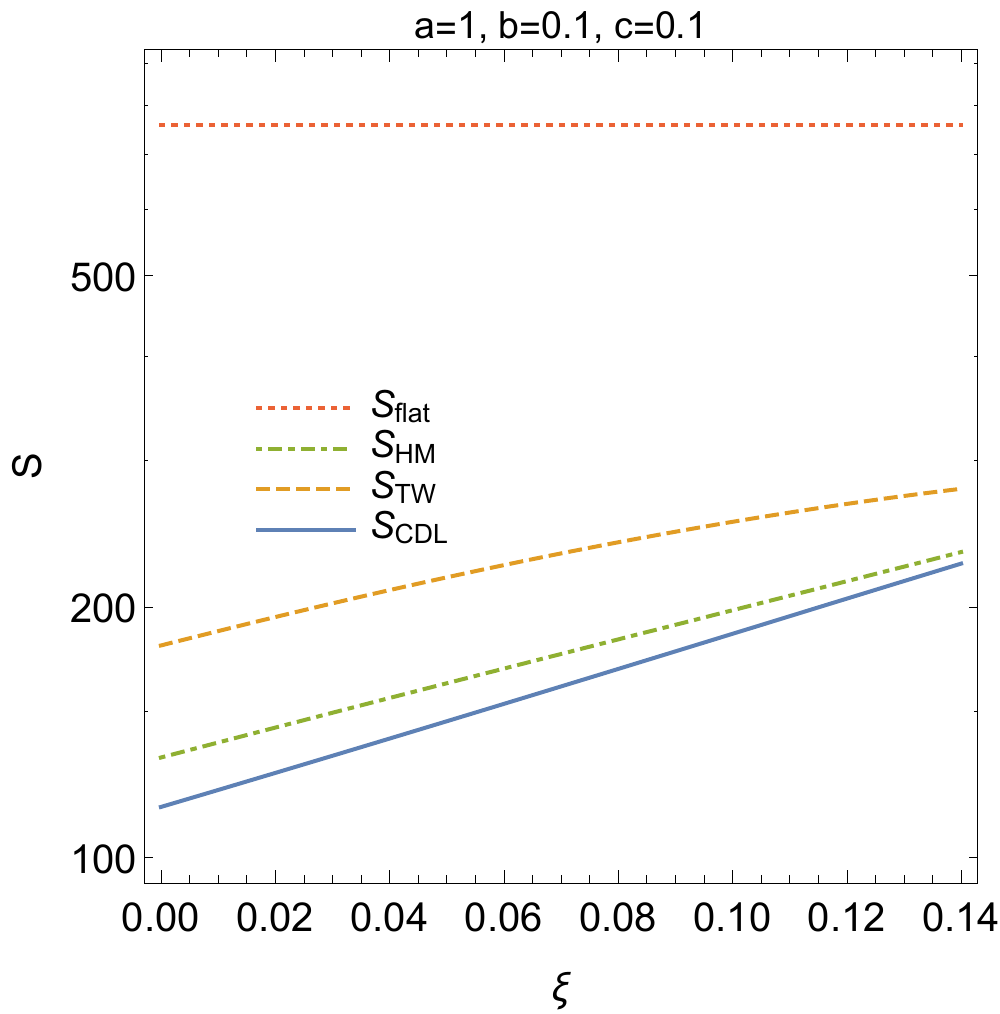}\\
\includegraphics[width=0.32\textwidth]{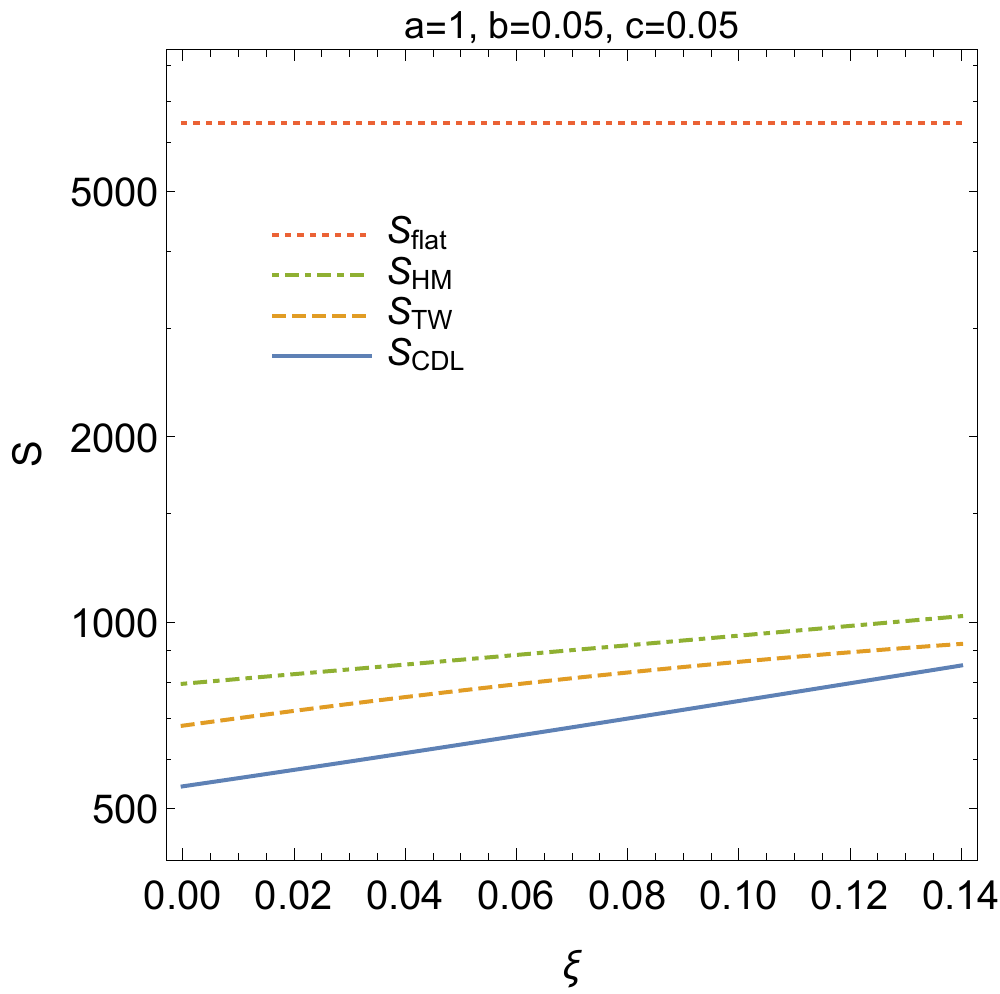}
\includegraphics[width=0.32\textwidth]{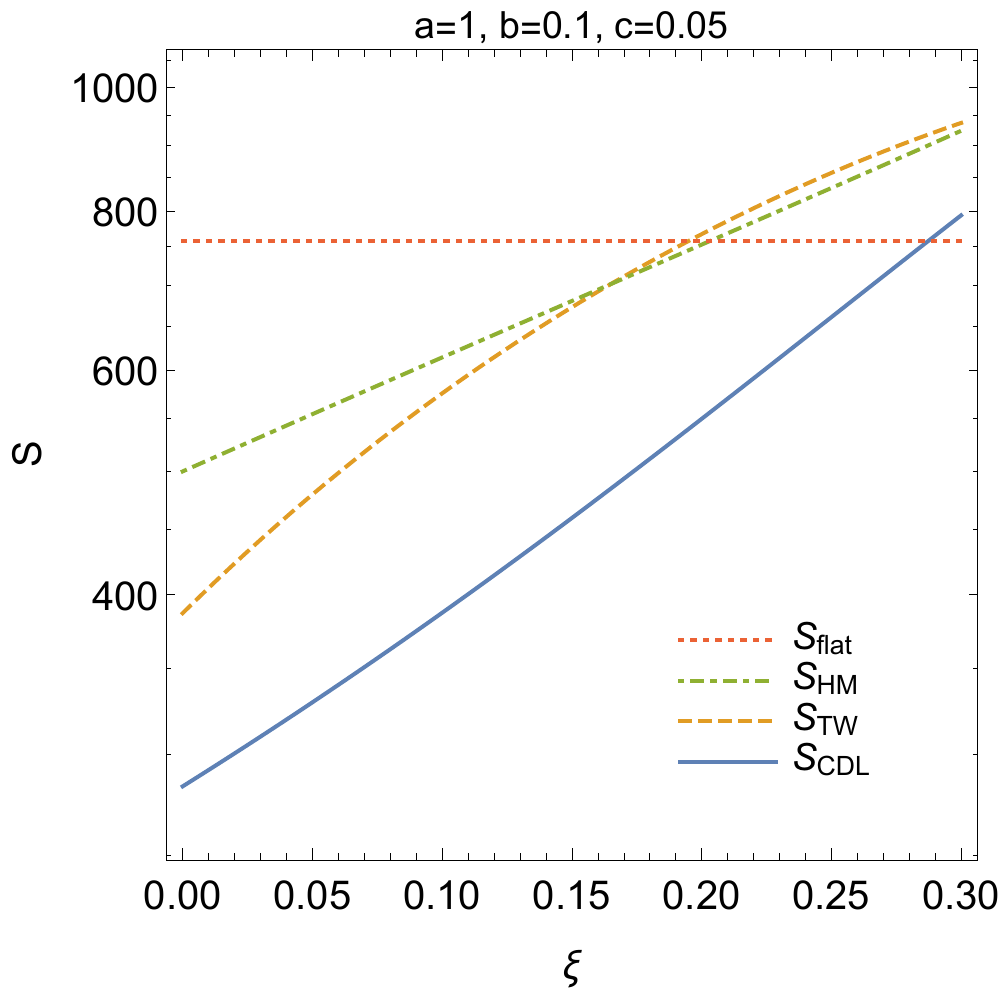}
\caption{ Tunneling action for de Sitter false vacuum as a function of non-minimal coupling obtained using four different methods. Parameter choice and labels are the same as in Figure \protect\ref{fig:actionMin}. Rows correspond to different false vacuum energy densities parametrised by $c=(0.1, \ 0.05)$ from top to bottom. \label{fig:actionGrav}}
\end{center}
\end{figure}

To calculate the action of our solution and consequently the tunneling probability we numerically perform the action integral (\ref{eqn:euclideanactionsimple}), knowing the CDL solution for $\phi(\tau)$ and $\rho(\tau)$, and put it in (\ref{eq:CDLaction}) together with the background action (\ref{eqn:backgroundaction}). Figures~\ref{fig:actionMin} and \ref{fig:actionGrav} show the resulting action for all four methods discussed in this paper - $S_{\rm CDL}$ is the numerically obtained result including Einstein-Hilbert term and non-minimal coupling to gravity, $S_{\rm TW}$ is the result of its thin-wall approximation, $S_{\rm HM}$ comes from Hawking-Moss solution and $S_{\rm flat}$ is the numerically obtained flat spacetime result completely neglecting gravity. 

Figures \ref{fig:actionMin} and \ref{fig:actionGrav} show that both approximations, TW and HM, always overestimate the action which includes the influence of gravity. But the validity of them is different - for relatively large vacuum energies HM solution gives action smaller than thin-wall and is a very good approximation while for smaller vacuum energy thin-wall approximation becomes better and the suppression of the action due to gravitational effects lowers. However, both approximations become less accurate as the vacuum energy decreases. In particular, in the Minkowski case gravitational effects suppress vacuum decay by increasing the action and the HM solution does not exist ($S_{\rm HM}$ would be infinite). At the same time TW severely overestimates the action due to non zero coupling $\xi$.

\begin{figure}[ht]
\begin{center}
\includegraphics[height=3.85cm]{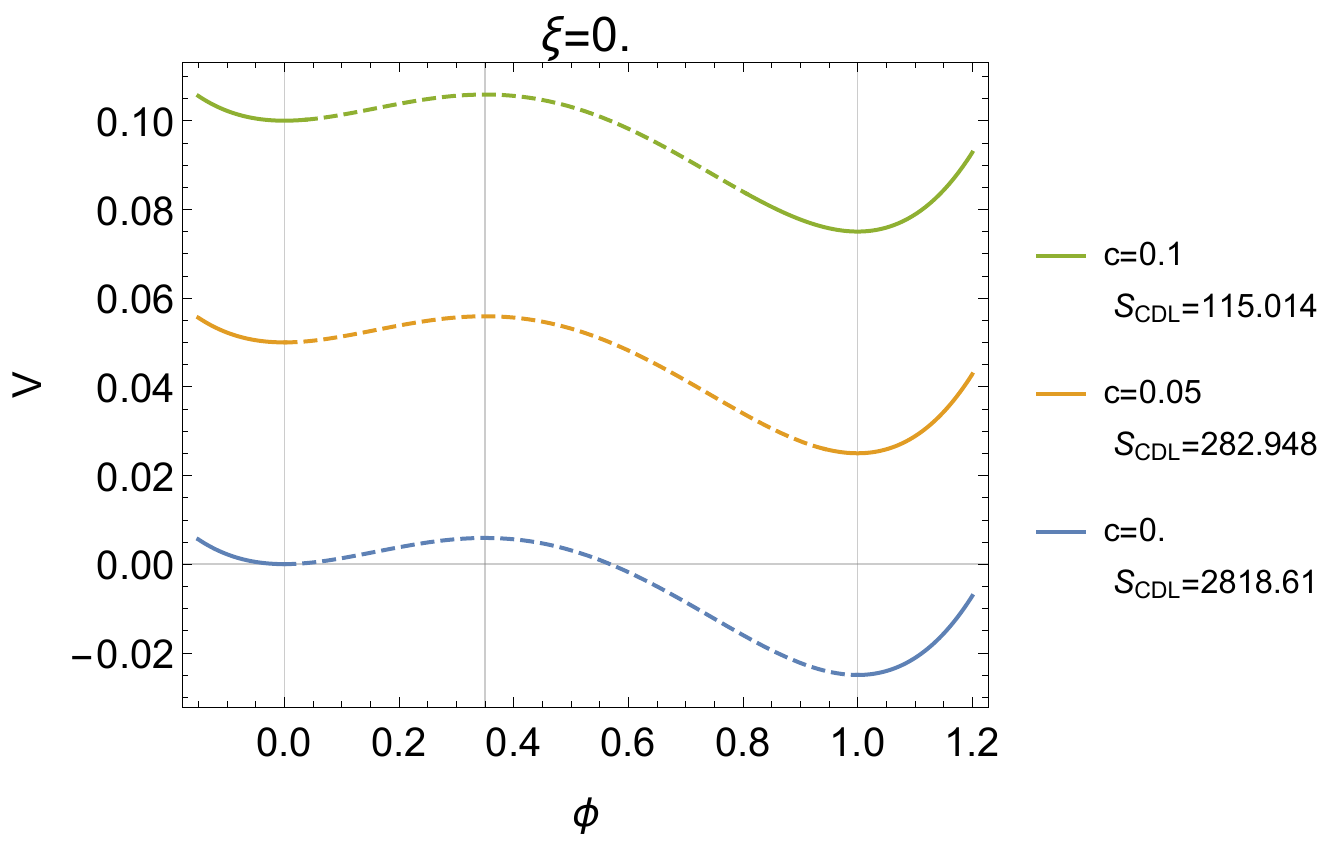}
\includegraphics[height=3.85cm]{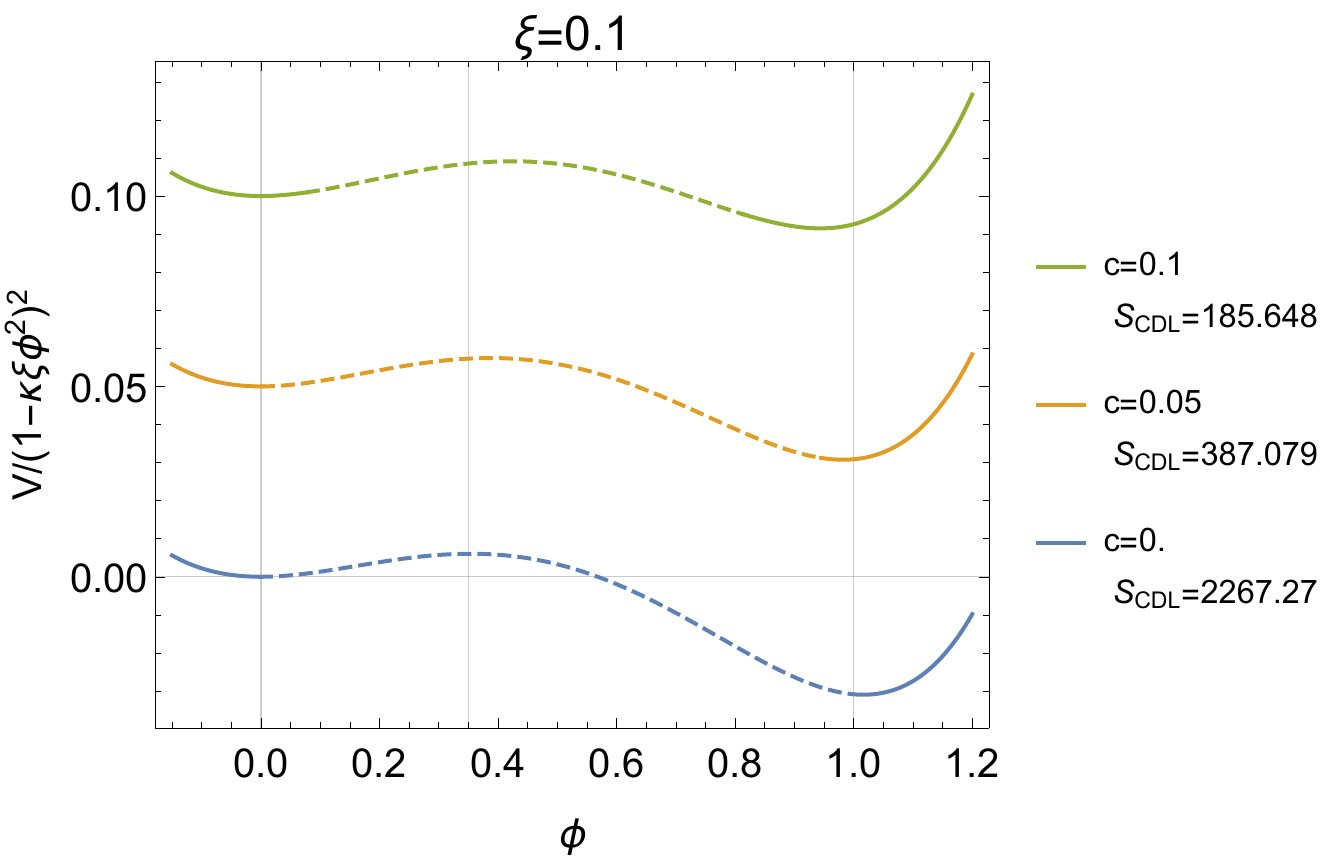}
\caption{ Potentials with different values of the vacuum energy $c$. The part of the potential actually probed by the tunneling solution is dashed. For this example the non-minimal coupling was set to $\xi=0$ (left panel) and $\xi=0.05$ (right panel) while the vacua splitting parameter $b=1/10$. \label{fig:thrmaltunneling}}
\end{center}
\end{figure}

Moreover, action quickly decreases as the false vacuum energy increases which can be interpreted as a result of a temperature effect coming from an effective temperature induced by our compact spacetime \cite{Brown:2007sd}. Presence of non-minimal coupling weakens this effect as it makes the potential more and more flat as the vacuum energy increases, thus also increasing the action. In this case bounce solutions do not have to reach the false vacuum exactly but they only have to pass the bubble wall. Figure~\ref{fig:thrmaltunneling} depicts the potentials with different values of the vacuum energy $c$ and part of the potential actually probed by the tunneling solution.

For a fixed positive vacuum energy (given $c$) increasing $\xi$ also results in more flat potential. It means that the bounce probes only values closer to the top of the barrier which makes it more similar to the HM solution. Furthermore, when value of $\xi$ is too large potential becomes too flat and CDL bounces cease to exist \cite{Hackworth:2004xb,Artymowski:2015mva}. Thus, as the vacuum energy decreases larger values of $\xi$ allow tunneling. 

\section{Conclusions and summary \label{sec:concl}}

We analysed the vacuum decay process in presence of non minimal coupling to gravity for a general but simple model consisting of a single neutral scalar with the generic potential described in Section~\ref{sec:model}. We developed a thin-wall solution concerning $\xi$ and provided explicite formulae needed to compute the decay exponent in a generic model. Analytical results were verified by a precise numerical calculation.

Influence of non-minimal coupling to gravity is very different in cases of Minkowski and dS false vacua. For dS vacuum decay probability quickly decreases
as the coupling grows and the vacuum can be made absolutely stable. In Minkowski case effect is much weaker and the decay rate increases for small values of $\xi$. Also, for a flat spacetime TW approximation works worse significantly overestimating tha action due to $\xi$ term. Even though TW approximation may not give a precise result in a specific model, at least the order of magnitude is right - especially in dS case where gravitational correction decreases the vacuum stability.

\section*{Acknowledgements}
This work has been supported by the Polish NCN grants DEC-2012/04/A/ST2/00099 and 2014/13/N/ST2/02712 and also partially supported by the University of Adelaide and the Australian Research Council through the ARC Centre of Excellence for Particle Physics at the Terascale (CoEPP) (CE110001104). ML and OC were also supported by the doctoral scholarships numbers 2015/16/T/ST2/00527 and 2016/20/T/ST2/00175, respectively. OC thanks Bonn Bethe Center for Theoretical Physics for hospitality.

\end{document}